\documentclass[a4paper,UKenglish,cleveref, autoref, thm-restate]{lipics-v2019}

\newcommand{\ra}{\rightarrow}

\title{Fast and Work-Optimal Parallel Algorithms for Predicate Detection}

\author{Rohan Garg}{Purdue University, Department of Computer Science}{rohanvgarg@gmail.com}{}{}{}

\authorrunning{Anonymous} 

\Copyright{Anonymous} 



\category{} 

\relatedversion{} 

\supplement{}

\nolinenumbers 

\usepackage{thmtools,thm-restate}
\usepackage{tikz}
\usepackage{makecell}
\usetikzlibrary{positioning}
\usetikzlibrary{decorations.pathreplacing}
\usepackage{makecell}

\EventEditors{John Q. Open and Joan R. Access}
\EventNoEds{2}
\EventLongTitle{42nd Conference on Very Important Topics (CVIT 2016)}
\EventShortTitle{CVIT 2016}
\EventAcronym{CVIT}
\EventYear{2016}
\EventDate{December 24--27, 2016}
\EventLocation{Little Whinging, United Kingdom}
\EventLogo{}
\SeriesVolume{42}
\ArticleNo{23}

\begin{document}
\maketitle








\begin{abstract}
Recently, the predicate detection problem was shown to be in the parallel complexity class \textit{NC}. In this paper, we give the first work-optimal parallel algorithm to solve the predicate detection problem on a distributed computation with $n$ processes and at most $m$ states per process. The previous best known parallel predicate detection algorithm, \textit{ParallelCut}, has time complexity $O(\log mn)$ and work complexity $O(m^3n^3\log mn)$. We give two algorithms, a deterministic algorithm with time complexity $O(mn)$ and work complexity $O(mn^2)$, and a randomized algorithm with time complexity $(mn)^{1/2 + o(1)}$ and work complexity $\tilde{O}(mn^2)$. 
Furthermore, our algorithms improve upon the space complexity of \textit{ParallelCut}. Both of our algorithms have space complexity $O(mn^2)$ whereas \textit{ParallelCut} has space complexity $O(m^2n^2)$.




\end{abstract}






\section{Introduction}
\label{S:1}

Ensuring the correctness of distributed systems and concurrent programs is a challenging task. A bug may appear in one execution of the system, corresponding to a particular thread schedule, but not in others. One of the fundamental problems in debugging these systems is to check if the user-specified condition exists in any global state of the system that can be reached by a different thread schedule. This problem, called predicate detection, takes a concurrent computation (in an online or offline fashion) and a condition that denotes a bug (for example, violation of a safety constraint), and outputs a schedule of threads that exhibits the bug if possible \cite{Garg02:bk, Marzullo-ConsGlobalStates}. Predicate detection is predictive because it generates inferred reachable global states from the computation; an inferred reachable global state might not be observed during the execution of the program, but is possible if the program is executed in a different thread interleaving.

The predicate detection problem has many applications. Many classic problems in distributed computing such as termination detection, deadlock detection, and mutual exclusion can be modeled as predicate detection. Similarly, classic problems in parallel computing such as mutual exclusion violation, data race detection, and atomicity violation can also be modeled as predicate detection. General predicate detection is NP-complete \cite{ChaGar:DC} and therefore researchers have explored special classes of predicates. In this paper, we present improvements to the work and space complexity of parallel algorithms for the class of conjunctive predicates. While there has been extensive work in online and offline distributed algorithms for conjunctive predicate detection, there is only one parallel algorithm in the literature for predicate detection called $ParallelCut$ \cite{Garg:2019}. It is shown in \cite{Garg:2019} that:\\

\textbf{Theorem (Garg and Garg \cite{Garg:2019})}: The conjunctive predicate detection problem on $n$ processes with at most $m$ states can be solved in $O(\log mn)$ time using $O(m^3n^3 \log mn)$ operations on the common CRCW PRAM.\\

Although this result places the predicate detection problem in the class $NC$, it has a very high work complexity of $O(m^3n^3 \log mn)$. Additionally, it has a space complexity of $O(m^2n^2)$. The high space complexity was required for \textit{ParallelCut} as the algorithm requires the transitive closure of a matrix to achieve its fast run time. These properties make this result impractical for adoption in practice. Our results reduce both the work complexity and space complexity of solving the conjunctive predicate detection problem. Both the algorithms presented in this paper have desirable work and space complexities that make them suitable for adoption in practice. A summary of our results' complexity measures is given in Table \ref{fig:table1}.

It should be noted that generally there are many more states along one process than there are processes in total. In essence, we should think of $m >> n$. Shaving off factors of $m$ from the work, space, and time complexities provides vast benefits in practice.


\begin{table}[h!]
\centering
\begin{tabular}{ |c|c|c|c| } 
 \hline
 Algorithm & Work & Time & Space \\ 
 \hline
 Sequential \cite{GargWald:WeakUnstable} & $O(mn^2)$ & $O(mn^2)$ & $O(mn^2)$\\
 \hline
 \textit{ParallelCut} \cite{Garg:2019} & $O(m^3n^3\log mn)$ & $O(\log mn)$ & $O(m^2n^2)$\\
 \hline
 \textbf{This Paper: }\textit{OptDetect} & $O(mn^2)$ & $O(mn)$ & $O(mn^2)$\\
 \hline
 \textbf{This Paper: }\textit{JLSDetect} & $\tilde{O}(mn^2) $ & $(mn)^{1/2 + o(1)}$ & $O(mn^2)$\\
 \hline
\end{tabular}
 \newline
\caption{Summary of previous results for conjunctive predicate detection.
\label{fig:table1}}

\end{table}

We give a work-optimal parallel algorithm, \textit{OptDetect}, for predicate detection. This is the first work-optimal parallel algorithm for predicate detection to the best of our knowledge. We show that the sequential algorithm presented in \cite{Garg:2019} can be parallelized and gives optimal work complexity bounds. Additionally, this algorithm can be used in an online fashion since it only looks at one state from each process per round. In the online setting, we assume that each process's state trace is loaded into a queue and in each round we can only look at the current head of each queue. This makes it particularly useful for periodic computations or infinite computations. The work complexity guarantee of $O(mn^2)$ matches both a lower bound on the number of operations and the sequential best shown in \cite{GargWald:WeakUnstable}. The lower bound argument is based on the number of comparisons it takes to find $n$ incomparable vectors in a given poset \cite{GargWald:WeakUnstable}. 

Our second result, is a fast parallel algorithm for solving the predicate detection problem. This fast algorithm, \textit{JLSDetect}, gives a better time complexity bound than the current sequential best\cite{GargWald:WeakUnstable} and a better work complexity bound than that of \textit{ParallelCut}.

In \textit{ParallelCut}, one of the steps is the solving of the single-source reachability problem. This problem is widely acknowledged to have a harsh time-work trade off \cite{KarpRama90}. Parallel reachability has been studied extensively in the literature and is known to have connections to long-standing open problems in complexity theory. Until the recent breakthrough of Fineman \cite{FinemanSTOC18}, all known parallel reachability algorithms with linear work had $O(|V|)$ time complexity where $V$ is the set of vertices in the directed graph.

The \textit{JLSDetect} algorithm is based off using a different reachability algorithm for this step. \textit{JLSDetect} is based on the parallel reachability algorithm of Jambulapati, Liu, and Sidford \cite{DBLP:conf/focs/LiuJS19}. 

In this paper, we contribute the two following theorems:\\
 

\textbf{Theorem 1}: The conjunctive predicate detection problem on $n$ processes with at most $m$ states can be solved by $OptDetect$ in $O(mn)$ time and $O(mn^2)$ space using $O(mn^2)$ operations on the common CRCW PRAM.
\\

\textbf{Theorem 2}: The conjunctive predicate detection problem on $n$ processes with at most $m$ states can be solved by $JLSDetect$ in $(mn)^{1/2 + o(1)}$ time and $O(mn^2)$ space using $\tilde{O}(mn^2)$ operations on the common CRCW PRAM with high probability in $mn$.
\\


\section{Our Model}
\label{sec:model}

We assume a message-passing system without shared memory or a global clock. A distributed system consists of a set of $n$ processes denoted by $P_1,P_2,...,P_n$ communicating via asynchronous messages. We assume that no messages are lost, altered or spuriously introduced. However, we do not make any assumptions about a FIFO nature of the channels.  In this paper, we run our computations on a single run of a distributed system. 

Each process $P_i$ in that run generates a single execution trace which
is a finite sequence of local states. The state of a process is defined by the values of all its variables including its program counter. 
Let $S$ be the set of all states in the computation.
We define the usual happened-before relation ($\ra$) on the states
(similar to Lamport's happened-before relation between events) as follows. 
If state $s$ occurs before $t$ in the same process, then $s \ra t$.
If the event following $s$ is a send of a message and the 	
event preceding $t$ is the receive of that message, then $s \ra t$. Finally,
if there is a state $u$ such that $s \ra u$ and $u \ra t$, then $s \ra t$.
A {\em computation} is simply the poset given by $(S, \ra)$.

\begin{figure}[htbp]
\begin{center}
\includegraphics[height=2.4in]{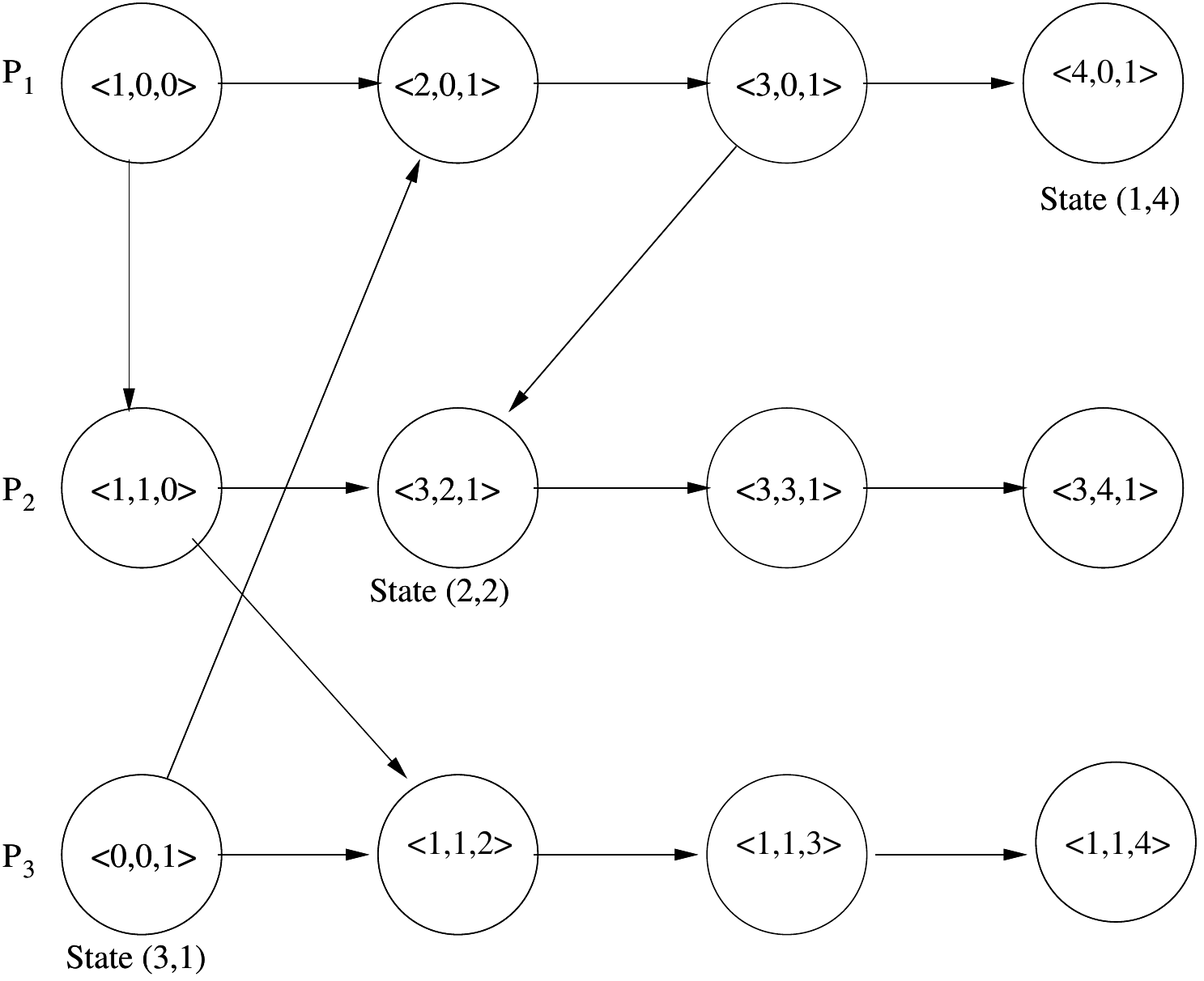}
\caption{\label{fig:model} State-Based Model of a Distributed Computation }
\end{center}
\end{figure}

It is helpful to define what a {\em consistent global state} of a computation is.
Let $s || t$ denote that states $s$ and $t$ are incomparable, i.e., $s || t \equiv  s \not \ra t \wedge t \not \ra  s$.
A {\em consistent global state} $G$ is an array of states such that
$G[i]$ is the state on $P_i$ and $G[i] || G[j]$ for all $i,j$.
Consistent global states model {\em possible} global states in a parallel or a distributed computation.
We assume that there is a vector clock algorithm \cite{Matt:VTime,Fidge:PartialOrders} running with the computation that tracks
the happened-before relation. A vector clock algorithm assigns a vector $s.v$ to every state $s$ such that 
$s \rightarrow t$ iff  $s.v < t.v$.
The vectors $s.v$ and $t.v$ are called the {\em vector clocks} at $s$ and $t$. 
Fig. \ref{fig:model} shows an example of an execution trace with vector clocks.


A {\em local predicate} is any boolean-valued formula on a local
state.  For example, the predicate ``$P_i$ is in the critical section'' is a local predicate.
It only depends on the state of $P_i$, and $P_i$ can obviously detect
that local predicate on its own.
A {\em global predicate} is a boolean-valued formula on a global state. 
For example, the predicate $(P_1~ \mbox{is in the critical section}) \wedge (P_2 ~ \mbox{is in the critical section})$ is a global predicate.
A global predicate depends upon the states of many processes.
Given a computation $(S, \ra)$, and a boolean predicate $B$, the {\em predicate detection} problem is to determine if there exists
a consistent global state $G$ in the computation such that $B$ evaluates to true on $G$. We restrict the input so that we only consider states for which the local predicate for that process evaluates to true. 

We focus on Weak Conjnctive Predicates (WCP) in this paper.
A global predicate formed only by the conjunction of local predicates is called a Weak Conjunctive Predicate (WCP) \cite{GargWald:WeakUnstable}, or simply, a conjunctive predicate. 
Thus, a global predicate $B$ is a conjunctive predicate if it can be written as
$l_1 \wedge l_2 \wedge \dots \wedge l_n$, where each $l_i$ is a predicate local to $P_i$.
We restrict our
consideration to conjunctive predicates because any boolean expression of local predicates can be detected using
an algorithm that detects conjunctive predicates as follows. We convert the boolean expression into 
its disjunctive normal form. Now each of the disjuncts is a pure conjunction of local predicates and
can be detected using a conjunctive predicate algorithm. This class of predicates models a large number of possible bugs.

\section{Detecting Conjunctive Predicates in Parallel}
\label{sec:conj}
In this section, we outline some key properties used in the intuition behind predicate detection algorithms. A conjunctive predicate $B$ is of the form $B = l_1 \wedge l_2 \wedge \dots \wedge l_n$.
To detect $B$, we need to determine if there exists a consistent global state $G$ such that
$B$ is true in $G$. Note that given a computation on $n$ processes each with $m$ states, there
can be as many as $m^n$ possible consistent global states. Therefore, a brute force approach of enumerating and
checking the condition $B$ for all consistent global states is not feasible.
Since $B$ is conjunctive, it is easy to show \cite{GargWald:WeakUnstable} that $B$ is true iff there exists a set of states $s_1, s_2, ..., s_n$ such that
(1) for all $i$, 
$s_i$ is a state on $P_i$, (2) for all $i$, $l_i$ is true on $s_i$ and (3) for all $i,j$: $s_i \| s_j$.
Any predicate detection algorithm will either output such local states or guarantee that it is not possible to find them in the
computation.
When the global predicate $B$ is true, there may be multiple $G$ such that $B$ holds in $G$.
For conjunctive predicates $B$, it is known that there is a unique minimum global state $G$ that satisfies $B$
whenever $B$ is true in a computation \cite{ChaGar:DC}. 
We are interested in algorithms that return the minimum  $G$ that satisfies $B$ since the minimum $G$ corresponds to the smallest
counter-example to a programmer's understanding.


\section{\textit{OptDetect}: A Work-Optimal Deterministic Algorithm for Predicate Detection}

\begin{figure}[htbp]
\fbox{
\begin{minipage} [b] {5.0in}
\begin{tabbing}
\=\=xxx\=xxx\=xx\=xx\=\=xx\=xx \kill
\> {\bf function} OptDetect()\\
\> Input: $states: array[1 \dots n][1 \dots m]$  of vectorClock; \\
\> // sequence of local states  given by vector clocks\\
\> Output: Consistent Global State as array $cut[1 \dots n]$ \\
\\
\> Step 1: Create $cut$: set of initial states \\
\>\> {\bf var} $cut: array[1 \dots n ]$ of vectorClock;\\
\>\> {\bf var} $current: array[1 \dots n]$ of $\{1 \dots m\}$\\
\> \>{\bf for} $i := 1$ {\bf to} $n$ in parallel {\bf do}\\
\>\>\> $current[i] := 1$ ;\\
\>\>\>$cut[i] := states[i,current[i]]$ ;\\
\\
\> Step 2: Create $color$: array[$1$..$n$] of $\{red, green\}$;\\
\>\> {\bf var} $color: array[1 \dots n ]$ of $\{red, green\}$ {\bf init} $green$ \\
\> \>{\bf for all} ($i \in 1 \dots n, j \in 1 \dots n)$ in parallel {\bf do}\\
\>\>\> {\bf if} $cut[i] \rightarrow cut[j]$ then\\
\>\>\>\> $color[i] := red$;\\
\\
\> Step 3: Advance $cut$ in parallel \\
    \>\>{\bf for all} $(i \in 1 \dots n)$ in parallel {\bf do}\\ 
    \>\>\>{\bf if} $color[i] = red$ {\bf then}\\
    \>\>\>\> {\bf if} $cut[i]$ is the last state on its process {\bf then}\\
    \> \> \>\> \> output("No satisfying  Consistent Cut"); \\
    \>\>\>\>{\bf else}  \\
    \> \>\>\>\> $current[i] := current[i] + 1$; \\
    \>\>\>\>\> $cut[i] := states[i, current[i]]$;\\
    \>\> \> \>\>$color[i]$ := $green$;\\
    \>\> \> \>\>{\bf for} $j := 1$ {\bf to} $n$ in parallel {\bf do}\\
   \> \> \> \> \>\>{\bf if} $(color[j] = green)$ {\bf then}\\
   \> \> \> \> \>\>\>{\bf if} $(cut[i] \rightarrow cut[j])$ {\bf then} $color[i] := red$;\\
    \>\> \> \> \>\>\>{\bf if} $(cut[j] \rightarrow cut[i])$ {\bf then} $color[j] := red$;\\
   \> \> \> \>\>{\bf endfor};\\
    \>\>{\bf endfor};\\
    \>\\
    \> return $ConsistentCut := cut$ ;\\
\end{tabbing}
  \caption{ The \textit{OptDetect} algorithm to find the first consistent cut.
  \label{fig:OptDetect}}
\end{minipage}

} 
\end{figure}

In this section, we give the \textit{OptDetect} algorithm. The \textit{OptDetect} algorithm is the parallelization of the serial conjunctive predicate detection algorithm, \textit{SequentialCut}, presented in \cite{Garg:2019}. 

At a high level, we initialize the cut to the set of first states on each process. After this, we see if this cut contains states that are all concurrent with each other. If this is the case, we are done. If some states happened-before other states in this cut, we advance along those processes in parallel. We then re-evaluate our current cut to see if all states are concurrent with each other. This procedure repeats until we arrive at the first consistent cut. 

Since \textit{OptDetect} is the paralleliztion of the \textit{SequentialCut} algorithm, correctness follows from \cite{Garg:2019}. What remains to be shown are the time and work bounds. 

In step one of \textit{OptDetect}, we declare and initialize \textit{cut} and \textit{current} in constant time using $n$ processors. The data structures \textit{cut} and \textit{current} will be used to hold the current global state of the system and the index we are currently at on each process respectively. Thus, step one takes $O(n)$ operations. In step two, we declare and initialize \textit{color} which will be used to identify which states are ``bad''. This tells  us that we must advance on the processes that these states lie on. Step two takes $O(n^2)$ work since we use $O(n^2)$ processors and executing this step takes $O(1)$ time. In step three, we advance \textit{cut} in parallel to the first satisfying consistent cut. To analyze the time and work for this step, let us see how we color the states. Once a state has been colored \textit{red}, we advance along that state and never consider it again. Since there are at most $mn$ states, we consider at most $mn$ states. Each time we consider a state to be in the first consistent cut, we have to make $O(n)$ comparisons to the other $n-1$ states in \textit{cut}. So, in total, this step takes $O(mn^2)$ work. In the worst case we must explore all states to find the first consistent cut, so the time complexity is $O(mn)$. It should be noted that we get a time complexity of $O(mn)$ only in the pathological case where we only reject one state every time we advance on a process. Often, this algorithm will perform much better than $O(mn)$. Since this algorithm has been broken down into steps, we just have to identify the steps that have the largest time and work complexities. In this case, step three has both the largest time and work complexity of $O(mn)$ and $O(mn^2)$ respectively. 

Notice that this algorithm can be used in an online fashion since it only looks at one state from each process per round. In the online setting, we assume that each process's state trace is loaded into a queue and in each round we can only look at the current head of each queue. This makes it useful for applications that require periodic computations or infinite computations.

Lastly, notice that we only ever use the \textit{states} input of vector clocks and no data structure larger than \textit{states}. So, the space complexity of \textit{OptDetect} is $O(mn^2)$ since there are at most $mn$ states and each is identified by a vector clock of size $O(n)$.

In Figure \ref{fig:OptDetect}, we give the 
\textit{OptDetect} algorithm that solves the predicate detection problem in $O(mn)$ time using $O(mn^2)$ operations on the common CRCW PRAM. 


\section{\textit{JLSDetect}: A Fast Randomized Algorithm for Predicate Detection}

\begin{figure}[htbp]
\begin{center}
\includegraphics[height=2.4in]{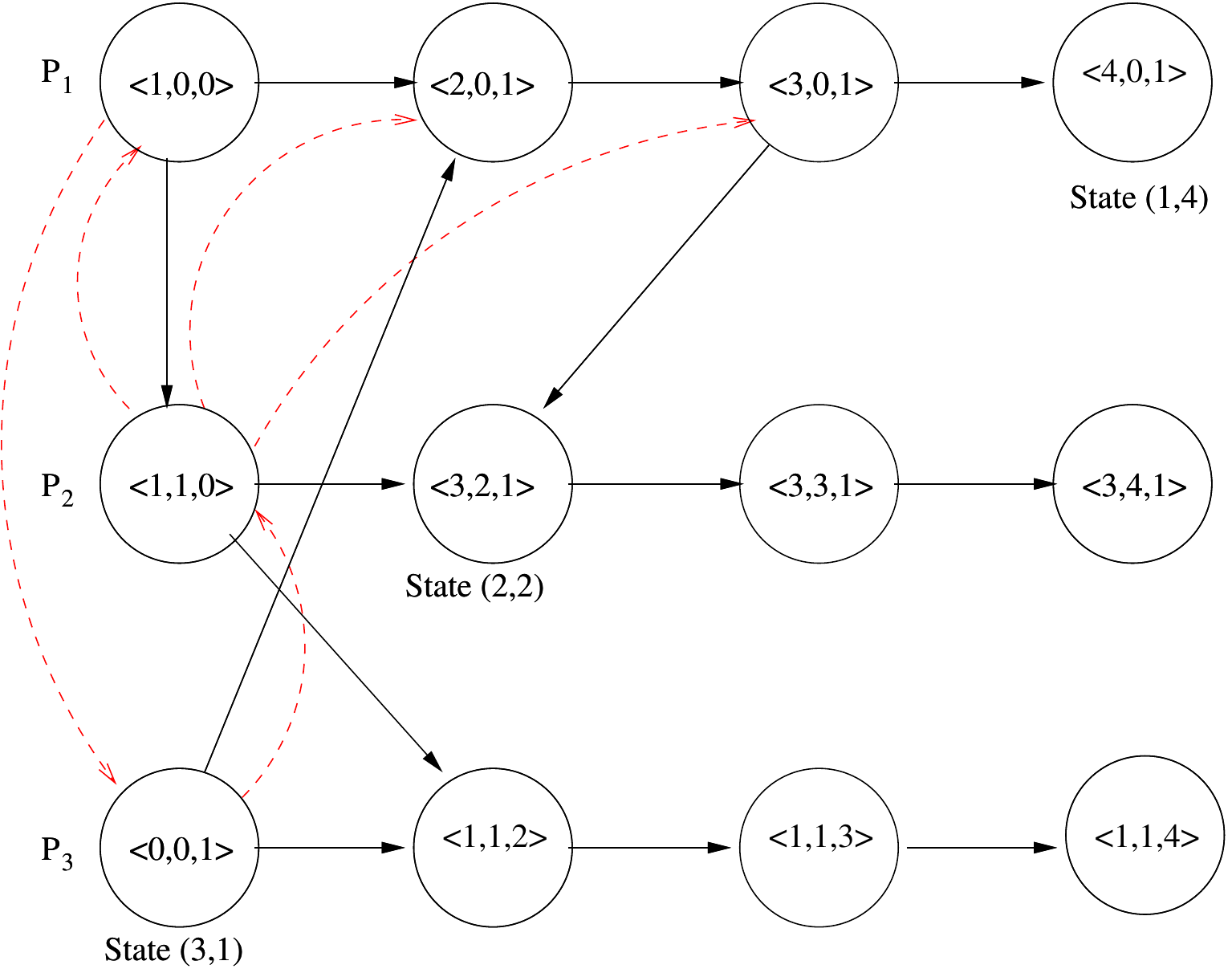}
\caption{\label{fig:reject} State Rejection Graph of a computation shown in dashed arrows }
\end{center}
\end{figure}

Next, we describe the second algorithm, \textit{JLSDetect}. \textit{JLSDetect} is based loosely around the idea of computing reachability of rejected states in the state rejection graph of a distributed system. This idea is presented in \cite{Garg:2019}. In Figure \ref{fig:reject}, we show what the state rejection graph looks like for the given computation in Figure \ref{fig:model}.

\textit{JLSDetect} improves upon the space complexity of \textit{ParallelCut} by using a smaller \textit{modified} data structure to represent to represent the state rejection graph. 
We are able to get a strong time complexity guarantee of $(mn)^{1/2 + o(1)}$. This is because we use the parallel reachability algorithm presented in \cite{DBLP:conf/focs/LiuJS19}. From here on, we will refer to this parallel reachability algorithm due to \cite{DBLP:conf/focs/LiuJS19} as \textit{JLSReach}.

At a high level, 
\textit{JLSReach} is actually taking in a graph with $|V|$ vertices and adding \textit{shortcut} edges to reduce the diameter to $|V|^{1/2 + o(1)}$ $w.h.p$. After this is done, the traditional \textit{ParallelBFS} can be utilized since \textit{ParallelBFS} runs in linear work and time proportional to the diameter of the graph. The output after reducing the diameter and running \textit{ParallelBFS} is the set of nodes reachable from a specified source node $s$ in the diameter-reduced graph $G$. For a more detailed explanation of the algorithm, we refer the reader to \cite{DBLP:conf/focs/LiuJS19}.


In \textit{ParallelCut}, the algorithm uses a state rejection graph of size $mn$ by $mn$ to compute the consistent cut. The state rejection matrix allows us to query two states $s$ and $s'$ and tells us if the rejection of state $s$ implies the rejection of state $s'$. A more detailed explanation of the state rejection graph is given in \cite{Garg:2019}.

To reduce the space complexity, we want to reduce the representation size of the state rejection graph. To do this, we exploit the structure of the predicate detection problem to create a modified incidence matrix that represents the state rejection graph. The state rejection graph, $R$, is now represented as a smaller incidence matrix instead of as an adjacency matrix. We call this incidence matrix the \textit{state-max incidence matrix}. The \textit{state-max incidence matrix} is smaller due to the following observation.

Notice that each state has at most $n$ edges to other states in the state rejection graph. Consider the state-based representation corresponding to a single run of a distributed system. If some state $s = (i,j)$ has multiple edges in the graph such that these edges point to states on the same processor, we only need to consider the edge that points to the state with largest $j$ value, say state $s' = (i',j')$. This is because the \textit{rejection} of state $s$ implies the rejection of state $s'$ but \textit{also} implies the rejection of \textit{all} states that happened-before state $s'$. Trivially, this includes all states that happened-before state $s'$ and are on the same process as $s'$.

This allows us to only use a matrix indexed by states on one axis and by processors on the other axis. Now, our \textit{state-max incidence matrix} is of size $O(mn^2)$. Instead of explicitly, writing out each state on one axis of the matrix, we can use pointers to point back to the corresponding states in the input \textit{states}. For each state, we keep track of the largest state along all other processors for which it has an edge pointing to that state. More formally, we populate $R$ as follows: 

\[ R[(i,j), i'] = j' \equiv
 ((i',j') \ra (i, j+1) \wedge (\nexists ~(i',j'') ~|~ (j'' > j) \wedge (i',j'') \ra (i, j+1)))
  \]
  
\begin{figure}[htbp]
\fbox{
\begin{minipage} [b] {5.0in}
\begin{tabbing}
\=x\=xxx\=xxx\=xx\=xx\= \kill

\> {\bf function} JLSDetect()\\
\> Input: $states: array[1 \dots n][1 \dots m]$  of vectorClock \\
\> // Sequence of local states at each process\\


\> Output: Consistent Global State as array $cut[1 \dots n]$ \\
\\

\> Step 1: Create $F$: set of states rejected in the first round\\
\>\> {\bf var} $F: array[1 \dots n ]$ of $ 0 \dots 1$  {\bf initially} $0$; 
\\
\>\> {\bf for all} $ (i \in 1 \dots n, j \in 1 \dots n)$ in parallel do\\
\>\>\>\>{\bf if} $((i,1) \ra (j,1)) $ {\bf then}\\
\>\>\>\>\> $F[i] := 1$;\\
\\

\> Step 2: Create $R$: State Rejection Graph\\
\>  // Represented as a State-Max Incidence Matrix \\
\>\>{\bf var} $R: [(1 \dots n,1 \dots m), (1 \dots n)]$ of $0 \dots m$\\
\>\> {\bf for all} $(i \in 1 \dots n, j \in 1 \dots m, i' \in 1 \dots n)$\\
\>\>\> $R[(i,j),i'] = 0$\\
\>\>{\bf for all} $ ( i \in 1 \dots n, j \in 1 \dots m) $ in parallel do\\
\>\>\> $R[(i,j),i] = j$; \\
\>\> {\bf for all} $(i \in 1 \dots n, j \in 1 \dots m-1)$ in parallel do\\
\>\>\>{\bf for} $(i' \in 1 \dots n)$ in parallel do\\
\>\>\>\>\> $loadVector = states[i][j+1];$\\
\>\>\>\>\> $R[(i,j),i'] = loadVector[i']$;\\
\\

\> Step 3: Create $RR$: set of nodes reachable from $F$ using $R$\\
\>\> {\bf var} $RR: array[(1 \dots mn)]$ of $0 \dots 1$ \\
\>\> $V(R) := V(R) \cup \{f\}$\\
\>\> $E(R) := E(R) \cup E_f$\\
\>\> $RR = JLSReach(R,f)$\\
\\

\> Step 4: Create $valid$: replace invalid states by $0$\\
\>\>{\bf var} $valid: array[[1 \dots n][1\dots m]$ of $ 0 \dots 1$; \\
\>\>{\bf for all} $( i \in 1 \dots n, j \in 1 \dots m)$ in parallel  do\\
\>\>\> $valid[i][j] := 1$;\\
\>\>{\bf for all} $( i \in 1 \dots n, i' \in 1 \dots n, j' \in 1 \dots m)$ in parallel  do\\
\>\>\>{\bf if} $( F[i] = 1 ) \wedge (RR[(i',j')] = 1)$ {\bf then} \\
\>\>\>\>$valid[i'][j'] := 0$; \\
\\

\\
\>Step 5: Create $cut$: First Consistent Global State \\
\>\> {\bf var} $cut: array[1 \dots n]$ of $0 \dots m $ {\bf initially} 0;\\ 
\>\> {\bf for all} $(i \in 1 \dots n, j \in 1 \dots m)$ in parallel do\\
\>\>\>\> {\bf if} $(valid [i][j] \neq 0)$ {\bf then}  \\
\>\>\>\> \>  {\bf if} $(j=1) \vee ((j>1) \wedge (valid [i][j-1]  = 0)$ {\bf then}  \\
\>\>\>\>\>\> $cut[i] := j$;\\
\>\> {\bf for all} $(i \in 1 \dots n)$ in parallel do\\
\>\>\>\> {\bf if} $(cut[i] = 0)$  {\bf then} \\
\>\>\>\>\> output("No satisfying  Consistent Cut");\\
 \\
\>    \> return $ConsistentCut := cut$;\\
\end{tabbing}
\caption{The JLSDetect algorithm to find the first consistent cut. \label{fig:JLSDetect}}
\end{minipage}
 } 

\end{figure}

Populating this \textit{state-max incidence matrix} can be done in $O(1)$ time since the relation we have described above is exactly what is stored in the vector clock representation of a state. If some state $s$ has vector clock $\Vec{v_s} = [ s_1, s_2, \dots , s_n]$, then by the definition of vector clock, $s_i$ is the largest process on process $i$ that happened-before $s$. So, to set the the \textit{state-max incidence matrix}, we will, for state $s = (i,j)$, load in the vector clock of state $s' = (i, j+1)$ into $R$ such that $R[(i,j), i'] = \Vec{{v_{s'}}}[i']$.  By using a separate processor for each state, and then $n$ processors to load in the vector clock, this can be done in $O(1)$ time and $O(mn^2)$ work.




The improved time complexity bound comes directly from the time complexity of computing reachability using \textit{JLSReach}.
For any $|V|$-node $|E|$-edge directed graph, \textit{JLSReach} computes all vertices reachable from a given source node $s$ with $\tilde{O}(|E|)$ work and $|V|^{1/2 + o(1)}$ time with high probability in $|V|$. Since we have at most $mn$ nodes in our state rejection graph and we know there are at most $mn^2$ edges in the state rejection graph, the work complexity and time complexity of using \textit{JLSReach} is $\tilde{O}(mn^2)$ and $(mn)^{1/2 + o(1)}$ respectively.

Now, we will informally explain the steps of the algorithm.

In step one, we create $F$, the set of all initially rejected states.  Let $I$ be the global state consisting of each processor's first local state, i.e., 
$ I = \{ (i,1) ~|~ i \in 1..n \} .$ If there are no dependencies between any of these states, we have already reached the first consistent global state. Else, if there is a dependency from one of these states to another, we reject whichever state happened-before the other and add it to $F$. We represent the set $F$ by a boolean bit array of size $n$ that is indexed by processor. This step can be done in $O(1)$ time in parallel with $O(n^2)$ work by using a separate processor for each value of $i$ and $j$.

In step two, we create the state rejection graph $R$ as a \textit{state-max incidence matrix} following the procedure given above.


In step three, we run \textit{JLSReach} on $R$. Notice that to use \textit{JLSReach} in step three, we must have a $single$ source node $s$. However, in our algorithm, $F$ may contain multiple nodes. Instead of running \textit{JLSReach} from each node in $F$, we can introduce a dummy source node $f$. We add the following set $E_f$ of edges to our state rejection graph:

$$ E_f = \{<f,v> ~|~ v \in F\}
$$

Now, we can run \textit{JLSReach} on $R$ from $f$ and this returns the set $RR$ of all vertices reachable from nodes in $F$ in the state rejection graph. In our given implementation, we use a Boolean bit array to represent set membership in $RR$. 

In step four, we mark which states are \textit{valid} by using both $F$ and $RR$. A \textit{valid} state is one that is part of a consistent cut. This step sets rejected states, or invalid states, with a $0$. This step can also be done in $O(1)$ time and $O(mn^{2})$ work.

Lastly, in step five,  we find the first consistent global state. To do this, we will look at our previous data structure \textit{valid}, and find the invalid state with the largest $j$ index along each process. Let this largest invalid state along process $p_i$ be $s = (i,j)$. Then, the first consistent global state contains the state $(i,j+1)$, namely, the state that appears right after state $s$. Of course, if the largest invalid state is the last state along a process, there does not exist a consistent global state. To compute the index of the largest invalid state, we will use a divide-and-conquer parallel reduce algorithm. For more on the parallel reduction operator, see \cite{DBLP:books/crc/99/BlellochM99}.

Let's treat the sequence of states along each process as a \textit{0-1} array where we have 0's as invalid states and 1's as valid states. To find the largest invalid state, i.e. the $0$-entry with the largest index, we will split the array into two halves and ask each half to compute its largest invalid state. Then, we give priority to the half that appears further on in the array. We continue splitting up these arrays until we are left with two entries. From here, we can simply compare the two entries and return index of the largest $0$. If neither entry has a $0$, we will return $0$. This process takes $O(\log m)$ rounds. Each round can be computed in $O(1)$ time if we have $O(mn)$ processors. So, this step takes $O(\log m)$ time and $O(mn)$ work. The recursive procedure to find this largest invalid state along each process is called function \textit{FLIS} and is given in Figure \ref{fig:FLIS}. \textit{FLIS} is used as a subroutine in \textit{JLSDetect}.

\begin{figure}[htbp]
\fbox{
\begin{minipage} [b] {5.0in}
\begin{tabbing}
\=\=xxx\=xxx\=xx\=xx\=\=xx\=xx \kill
\> {\bf function} FLIS()\\
\> Input: $S: array[1 \dots m]$  of $0 \dots 1$, \textit{start} index, \textit{end} index; \\
\> // Sequence of local states' validity given by $0$-$1$\\
\> Output: Index of largest 0 entry in $p_i$-$states$ \\
\\
\>\>{\bf if} $(end$ - $start == 1)$\\
\>\>\>{\bf if}$(S[start] == 0 \wedge S[end] == 1)~$ return $start$; \\
\>\>\>{\bf if}$(S[start] == 0 \wedge S[end] == 0)~$ return $end$; \\
\>\>\>{\bf if}$(S[start] == 1 \wedge S[end] == 0)~$ return $end$; \\
\>\>\>{\bf if}$(S[start] == 1 \wedge S[end] == 1)~$ return $0$; \\
\>\>{\bf else}\\
\>\>\>return $max(FLIS(S,start, end/2), FLIS(S,1+ end/2,end))$;
\end{tabbing}
  \caption{ The \textit{FLIS} subroutine to find the first the largest invalid state along a process $p_i$.
  \label{fig:FLIS}}
\end{minipage}

} 
\end{figure}

Now, that we have the largest invalid state along each process, we can compute the first consistent global state by taking the successor of each of these states. This takes $O(1)$ time with $O(n)$ processors.




Computing the time and work complexity of this algorithm boils down to finding the individual steps with the highest time and work costs. In this case, step three takes the longest time with a time cost of $(mn)^{1/2 + o(1)}$. The step with the largest work cost is also step three with a work cost of $\tilde{O}(mn^2)$ where $\tilde{O}(g(n))$ hides polylogarithmic factors in the function $g(n)$. Notice that the work-complexity is only off of the optimal by a $polylog(mn)$ factor. Similar to \textit{OptDetect}, the correctness of this algorithm follows from \cite{Garg:2019} since we have only replaced one reachability method with another. It is important to note that \textit{JLSDetect} is a randomized algorithm. \textit{JLSDetect} is given in Figure \ref{fig:JLSDetect}.

\section{Conclusions and Future Work}

We have given two algorithms which improve upon the best known parallel predicate detection algorithms in terms of work complexity and space complexity. We give the first work-optimal parallel algorithm for conjunctive predicate detection. Additionally, we give a fast randomized parallel predicate detection algorithm that outperforms the current sequential best and offers good work complexity guarantees.
Both the algorithms presented in this paper have properties that make them suitable for adoption in practice.

Recently, it was shown that many classical combinatorial optimization problems such as the stable marriage problem, market clearing price problem, and shortest-path problem can be cast as searching for an element that satisfies an appropriate predicate in a distributive lattice \cite{GargLLP-SPAA20}. Finding quicker predicate detection algorithms with strong work complexity guarantees is an open work with applications to this framework and potentially many other classic optimization problems. 

An open question that remains is: does there exist a parallel conjunctive predicate detection algorithm that runs in $O(polylog(mn))$ time using $O(mn^2)$ operations? An algorithm with these properties would lie in the parallel complexity class \textit{NC} and be work-optimal.  \\


\section{Acknowledgements}

The author would like to thank Vijay Garg for many helpful comments and discussions on this work.







\bibliography{sample2.bib}







\end{document}